# COVID-19 severity determinants inferred through ecological and epidemiological modeling

**Sofija Markovic[1], Andjela Rodic[1], Igor Salom[2], Ognjen Milicevic[3], Magdalena Djordjevic[2], Marko Djordjevic[1,*]**

[1]Quantitative Biology Group, Faculty of Biology, University of Belgrade, Serbia

[2]Institute of Physics Belgrade, National Institute of the Republic of Serbia, University of Belgrade, Serbia

[3]Department for Medical Statistics and Informatics, School of Medicine, University of Belgrade, Serbia

**Abstract**

Determinants of COVID-19 clinical severity are commonly assessed by transverse or longitudinal studies of the fatality counts. However, the fatality counts depend both on disease clinical severity and transmissibility, as more infected also lead to more deaths. Moreover, fatality counts (and related measures such as Case Fatality Rate) are dynamic quantities, as they appear with a delay to infections, while different geographic regions generally belong to different points on the epidemics curve. Instead, we use epidemiological modeling to propose a disease severity measure, which accounts for the underlying disease dynamics. The measure corresponds to the ratio of population averaged mortality and recovery rates ($m/r$). It is independent of the disease transmission dynamics (i.e., the basic reproduction number) and has a direct mechanistic interpretation. We use this measure to assess demographic, medical, meteorological and environmental factors associated with the disease severity. For this, we employ an ecological regression study design and analyze different US states during the first disease outbreak. Principal Component Analysis, followed by univariate and multivariate analyses based on machine learning techniques, is used for selecting important predictors. Without using prior knowledge from clinical studies, we recover significant predictors known to influence disease severity, in particular age, chronic diseases, and racial factors. Additionally, we identify long-term pollution exposure and population density as not widely recognized (though for the pollution previously hypothesized) predictors of the disease severity. We do not select meteorological factors as significant predictors of COVID-19 severity, though we previously found them to be significantly associated with the disease transmissibility. Overall, the proposed measure is useful for inferring severity determinants of COVID-19 and other infectious diseases, and the obtained results may aid a better understanding of COVID-19 risks.

**Introduction**

COVID-19 has brought large changes to people's lives, including significant impacts on health and the economy. COVID-19 effects (and those of other infectious diseases) at the population level can be characterized through the disease transmissibility and clinical severity. Transmissibility relates to the number of infected people, which in epidemiological models (see e.g. [1]) is quantified by the reproduction number $R(t)$ (corresponding to an average number of people infected by an individual during its infectious period). Clinical severity corresponds to the medical complications experienced by infected individuals, potentially also including death. In the epidemic models, two (population average) rates relate with the disease severity (see e.g. [2]): *i*) mortality rate ($m$) corresponding to the population-averaged probability per day that the detected case results in death, *ii*) recovery rate ($r$) corresponding to the inverse (population-averaged) time needed for a detected case to recover.

COVID-19 transmissibility and severity are often assessed through the numbers of confirmed cases and fatalities, respectively [3–8]. Regarding severity, a major complication is that the fatalities are correlated with infected numbers, as more infections leads to more fatalities. Additional complications are related to nonlinearities and delays that inherently characterize the disease dynamics. For example, deaths happen with a significant delay to infections, while number of fatalities in different regions (at a given time) may correspond to different points of the infected curve. Some of these problems can be alleviated by introducing corrections such as delay-adjusted case fatality rate (aCFR) [9–11], but their mechanistic interpretation is unclear [12]. Alternatively, we will here propose a relevant quantity with a clear mechanistic interpretation directly from epidemic modeling and derive how to infer that quantity from the available data. In particular, we will argue that the ratio of mortality and recovery rates (m/r) is a highly plausible population-level measure of disease severity: Higher mortality and lower recovery rates indicate a more severe disease leading to a larger m/r. We will also show (both theoretically and from empirical data) that this measure is a priori unrelated to R(t), which is a result independent from the specific assumed transmission mechanism.

To assess how reasonable is the proposed measure, it is desirable to use it to infer significant predictors (and their importance) of COVID-19 severity. However, this entails certain methodological challenges [13]. Specifically, significant predictors have to be selected among a large number of potentially relevant variables. Moreover, these variables may be mutually highly correlated [14,15], while interactions between them (and nonlinear relations) may also be relevant. To address this, we here use, to our knowledge, a unique approach for COVID-19, which combines Principal Component Analysis (PCA) and machine learning regression methods [16]. We will use linear regression with regularization and feature selection (allowing selecting significant predictors) and nonparametric methods based on ensembles of decision trees (that can accommodate interactions and nonlinear dependencies).

More generally, there has been some debate over applying ecological (top-down) vs epidemiological (bottom-up) models [17]. It was argued that a combination of these two may be optimal. The work presented here may be an example of this, exploiting the utility of both approaches. Specifically, epidemiological modeling will be used to propose an appropriate disease severity measure with a clear mechanistic interpretation. Ecological regression analysis will then be used to test the plausibility of this variable and reveal potential additional dependencies that may be hard to obtain from clinical studies.

## Methods

To extract the severity variable *m/r*, we used a modification of SEIR based compartmental model [2], introduced in our earlier paper [18]. The scheme of this (SPEIRD) model is presented in Figure 1. Note that *m/r* derivation is independent of the transmission mechanism and is (by construction) independent from the reproduction number *R(t)*. Consequently, the left rectangle (from which *R(t)* and its special case at the early stages of the epidemic, i.e., basic reproduction number ($R_0$), is determined) is presented only for clarity and coherence. The relevant part of the model represents the transition of the active cases (*A*) to healed (*H*) at recovery rate *r*, or to fatalities (*F*) at mortality rate *m*. Note that the cumulative (total) number of detected cases (*D*) corresponds to the sum of *A*, *H*, and *F*.

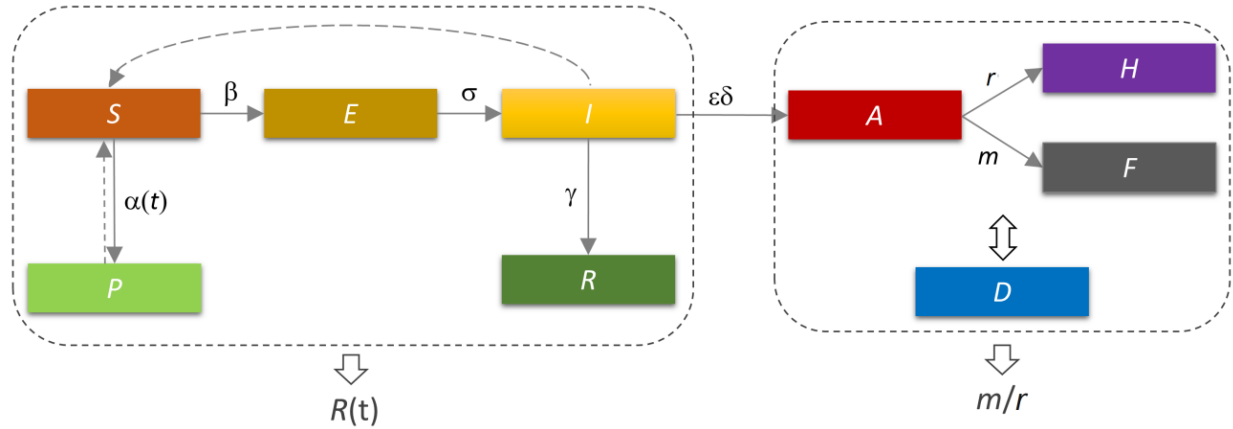

**Figure 1**. **Deriving the severity measure *m/r* from the epidemics compartmental model.** SPEIRD model is schematically shown. Transitions between the compartments are denoted by solid arrows, with the transition rates indicated above arrows. The dashed arrow from I to S indicates the interaction of I and S (infections) leading to the transition to E. The dashed arrow from P to S indicates the potential (reverse) transition from P to S due to the easing of measures. The dashed rectangles indicate parts of the model corresponding to the disease transmission (the left rectangle) and the disease outcome for the detected cases (the right rectangle). The single arrows indicate parts of the model from which the reproduction number R(t) and the severity measure (*m/r*) are, respectively, inferred. The total number of detected cases (D) corresponds to the sum of A, H and F and is denoted by a double arrow. Compartments are S – susceptible, P –protected, E – exposed, I –infected, R – recovered, A – active, H – healed, F – fatalities, D – total number of detected cases. *r* and *m* represent recovery and mortality rates of active (detected) cases.

The system of differential equations, which mathematically represents the model in Fig. 1 is given in [18]. From equations (5-6) in that paper, we obtain:

$$\frac{dH}{dt} = r \cdot A; \quad \frac{dF}{dt} = m \cdot A \Rightarrow \frac{dF}{dt} = \frac{m}{r} \cdot \frac{dH}{dt} \tag{1}$$

We integrate the right side of Eq. (1) from the epidemics start (t = 0) to the end (t = ∞):

$$F(\infty) = \frac{m}{r} H(\infty). \tag{2}$$

Since $D(t) = A(t) + F(t) + H(t)$, and since there are no more active cases at t = ∞, while $F(\infty)$ and $H(\infty)$ reach constant values (see Fig. 2A), we obtain:

$$D(\infty) = F(\infty) + H(\infty) \tag{3}$$

Combining Eqs. (2) and (3) gives:

$$\frac{m}{r} = \frac{CFR(\infty)}{1 - CFR(\infty)}; \quad CFR = \frac{F(\infty)}{D(\infty)}, \tag{5}$$

where CFR(∞) is the case fatality rate at the end of the epidemic. As the COVID-19 pandemic is still ongoing, we use the end of the first peak, where the number of active cases can be approximately considered as zero.

For consistency and easier direct comparison with the COVID-19 transmissibility analysis, data collection, data processing, and machine learning techniques are similar to the one presented in [19]. For completeness, full information is also provided in the Supplementary Methods, which also includes definition for all variables and principal components (PCs) used in the analysis. Supplementary Table contains all input data.

**Results**

Figure 2A illustrates inferring $m/r$ values. The cumulative number of detected cases and fatalities during the first peak of the epidemic is presented for one of the USA states (Connecticut). $m/r$ is inferred once both classes of the case counts reach saturation, leading to constant $m/r$ (inset in the figure). Figures 2B-C argue that $m/r$ is an independent observable of COVID-19 spread. A very high positive correlation (R = 0.97) between the cumulative number of fatalities and detected cases at a fixed time cross-section is obtained (Fig. 2B), quantitatively confirming the intuitive expectation that a higher number of infected is strongly related to higher fatality counts. On the other hand, the moderate correlation between $m/r$ and $R_0$ (Fig. 2C) is consistent with the a priori independence of these two variables. This moderate correlation reflects a genuine similarity in COVID-19 transmissibility and severity determinants (e.g., air pollution or weak immunity can be associated with both increased transmissibility [19] and severity of the disease [20]). Consequently, studies in which detected cases and fatalities are used as measures of, respectively, transmissibility and severity/mortality, strongly bias severity determinants towards those of transmissibility. This bias is resolved through $R_0$ and (here proposed) $m/r$ variables.

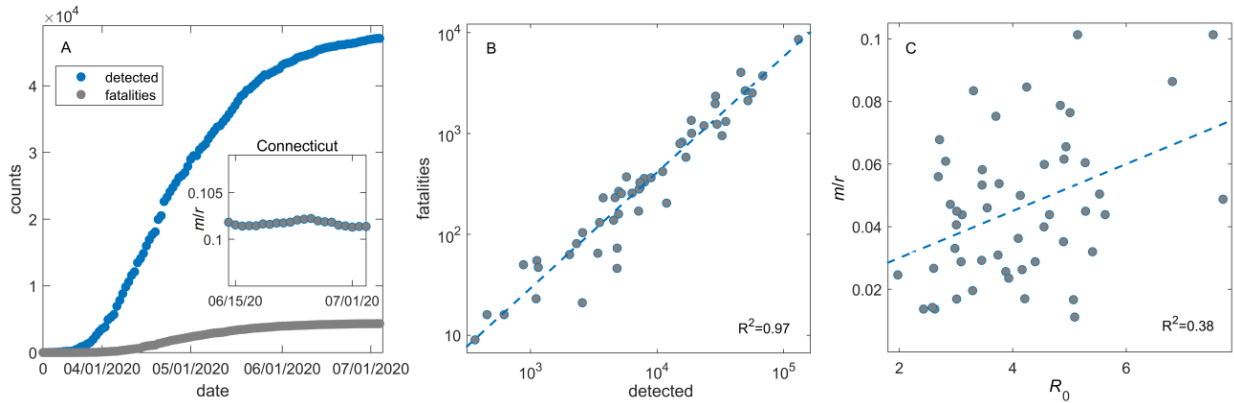

**Figure 2. Inferring $m/r$ from data.** A) Cumulative detected (D) and fatality (F) counts in Connecticut. $m/r$ is inferred from the time period (enlarged in the inset) corresponding to saturation (end of the first peak). B) and C) Correlation plots of F vs. D and $m/r$ vs R0 with the Pearson correlation coefficients shown.

Univariate analysis of $m/r$ relation to the variables used in the study is presented in Fig. 3. There are statistically significant correlations (P<0.05) of $m/r$ with several variables/PCs, as shown in Figure 3A and scatterplots (Figs. 3B-E). The highest (positive) correlation was observed for NO PC1, Disease PC4, and Density PC1, while the percentage of the youth population showed the highest negative correlation with $m/r$. Several other predictors, specifically, Density PC2, Disease PC2, $SO_2$, and NO Insurance PC1, Black, and PM2.5 also exhibit statistically significant correlations with $m/r$. As expected, chronic disease, pollution, population-density-related variables promote COVID-19 severity (positive correlations), as does the percentage of Afro-Americans (Black). Under 18 population percentage (Youth) decreases the severity (negative correlation), also as expected. Sign of the correlation with No Insurance PC1 is opposite than expected, as people with health insurance should get better medical treatment (further analyzed below).

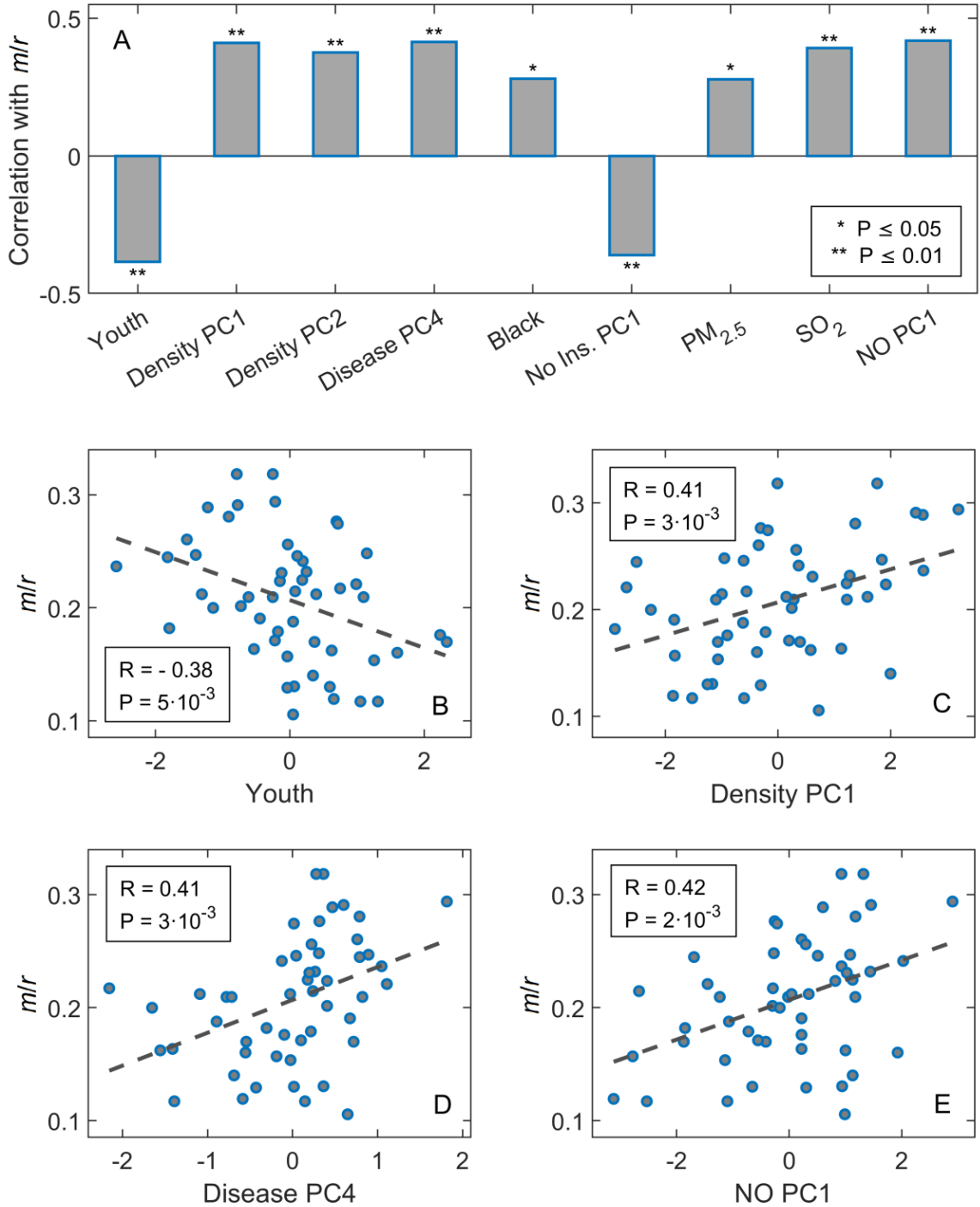

**Figure 3. Univariate correlation analysis.** (A) Values of Pearson's correlations for the variables significantly correlated (P<0.05) with $m/r$. Correlation plots of $m/r$ with (B) Youth (percent of the population under 18), (C) density PC1, (D) disease PC4, (E) NO PC1.

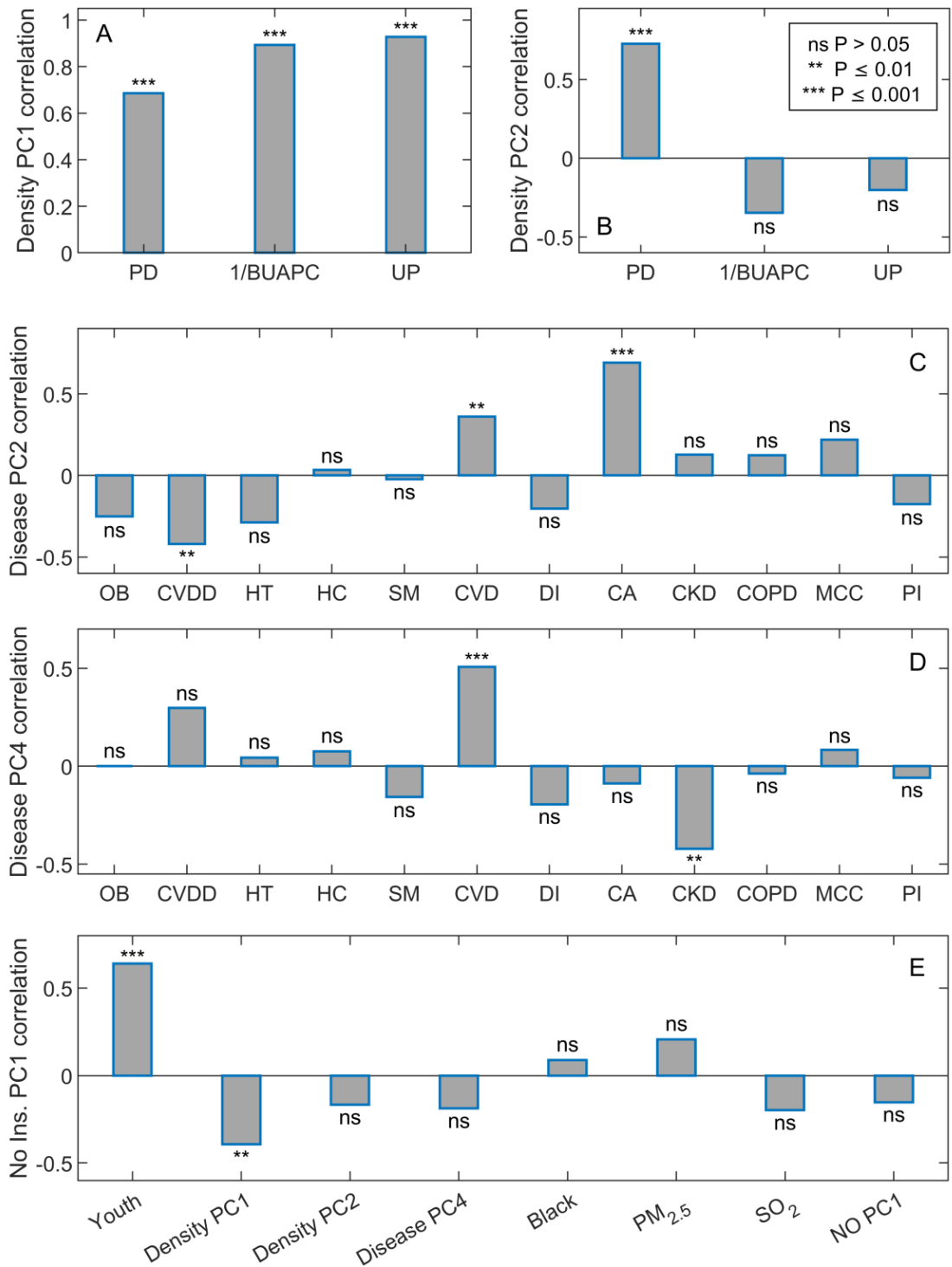

**Figure 4. Interpretation of the relevant PCs.** A) and B) Correlation of Density PC1 and Density PC2 with three population density variables; PD – population density, BUAPC – Built-Up Area Per Capita, UP – Urban Population, C) and D) Correlation of Disease PC2 and Disease PC4 with the variables from chronic disease group. OB – obesity, CVDD – cardiovascular disease deaths, HT – hypertension, HC – high cholesterol, SM – smoking, CVD – cardiovascular disease, DI – diabetes, CA – cancer, CKD – chronic kidney disease, COPD – chronic obstructive pulmonary disease, MCC – multiple chronic conditions, PI – physical inactivity, (E) Correlations of No Insurance PC1 with the variables from Fig. 3A.

Figures 4A-D provide interpretation of the relevant PCs by showing their correlations with the variables entering PCA. Density PC1 is comprised of all three parameters from the population density group (Figure 4A), presenting a general measure of population density, while Density PC2 is significantly correlated only with population density (Figure 4B). Disease PC2 and PC4 show, respectively, the highest positive correlation with the prevalence of cancer and cardiovascular diseases. Figure 4E shows a high correlation of No Insurance PC1 with Youth and Density PC1. Signs of these correlations, and the effect of these two variables on $m/r$, indicate that the unintuitive sign of No Insurance PC1 correlation with $m/r$ (noted above) is due to its high correlations with Youth and Density PC1.

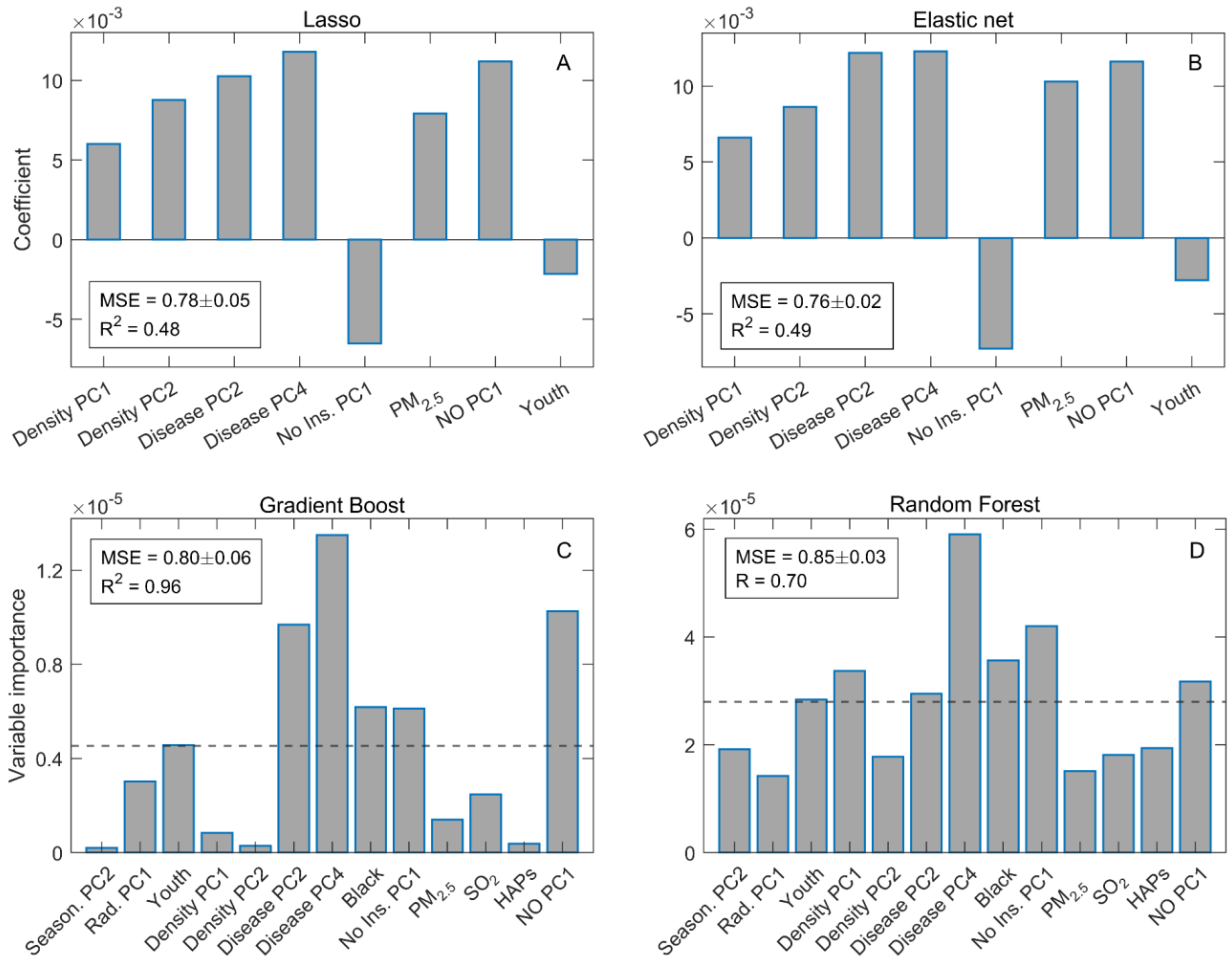

**Figure 5. Multivariate (machine learning) analysis.** Values of regression coefficients in relaxed A) Lasso and B) Elastic Net regressions. Only the variables whose coefficients are not shrunk to zero by the regressions are shown. The bars' height corresponds to the coefficients' value. Variable importance in C) Gradient Boosting and D) Random Forest regressions, with the bars' height corresponding to estimated importance. Testing set MSE values with the standard errors are shown for each model, corresponding to 5-fold cross-validations with 40 repartitions. Coefficients of determination on the entire dataset ($R^2$) are also shown for each model. Variable names are indicated on the horizontal axis.

We next perform multivariate analyses where the effect of each variable on $m/r$ is controlled by the presence of all other variables. Lasso and Elastic net also provide both regularization and the ability to single out significant predictors by shrinking the coefficients of other predictors to zero, i.e., to perform feature selection. This allows removing variables that do not affect m/r and would otherwise introduce noise in the model and likely result in high variance (overfitting). To eliminate the effect of noise on the estimated coefficient, which provides measures of the relative importance of the predictors, we employ relaxed Lasso and Elastic net procedures as described in Supplementary Methods. Both approaches robustly show similar results (Figs. 5A-B) and prediction accuracy (MSE indicated in figures). Disease PC4 appears in regressions as the most important predictor, followed by NO PC1 and Disease PC2. Other selected predictors are Density PC1 and PC2, No Insurance PC1, PM2.5, and Youth. These results agree with pairwise correlations, except for SO2 and Black, which appeared significant in pairwise correlation but were not selected by either linear of the regularization-based methods.

Next, we apply methods based on ensembles of decision trees, Gradient Boost and Random Forest (see Supplementary Methods). These are non-parametric machine learning methods, i.e., account for potentially highly non-linear relations and interactions between the predictors. For each of these methods, the predictor importance is presented in Figs. 5C-D, where the dashed lines indicate a standard threshold for identifying important predictors. Largely consistent results are obtained by both methods, where the predictors with the highest importance are Disease PC4, NO PC1, Disease PC2, No Insurance PC1, Blacks, and Youth. The only difference is in Density PC1, which appears as important in Random Forest but not in Gradient Boost. Results of Gradient Boost and Random Forest are also consistent with Lasso and Elastic Net, with an exception in Black (important in non-linear, but not linear, methods) and PM2.5 (vice versa). The effect of Black on m/r is therefore likely non-linear and/or involves interactions with other predictors (further discussed below).

To test our assumption that No Insurance PC1 appears in regressions due to its high correlation to other *m/r* predictors (mainly Youth and Density PC1), we next repeated the analysis, this time excluding No Insurance PC1. The results presented in Supplementary Figure S1 show that removing No Insurance PC1, besides leading to an (expected) increase of importance of Youth and Density PC1, does not significantly alter previously obtained results. Besides our assumption, this also confirms the robustness of the computational procedure.

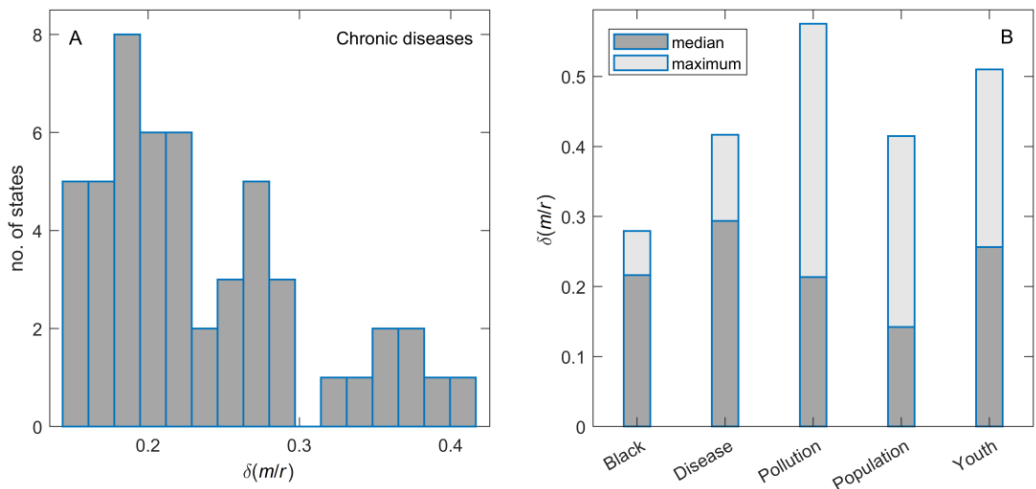

**Figure 6. Estimated change in m/r due to variations of important predictors**. A) Distribution of relative changes in *m/r* (δ(*m/r*)) due to variations in prevalence of chronic diseases observed in USA states. For each state, m/r was predicted for the range of the disease prevalence values observed throughout all other states. B) The same as in A) is repeated, but for the groups of predictors indicated on the horizontal axis. For each group, the median and maximal value of δ(*m/r*) is reported. δ(*m/r*) values for each group of predictors are estimated as described in Supplement Methods.

Finally, in Figure 6, we quantitatively estimate the influence of the five most important predictors determined above. For each of 51 states, we fix the values of all other predictors while changing the analyzed predictor's value within the range observed in all other provinces. The resulting distribution of the relative changes in m/r ($\delta(m/r)$) due to the variation of Chronic disease is shown in Fig. 6A, where each data point in the distribution corresponds to a single USA state. One can observe that changing Chronic disease values in a realistic range leads to significant variations of m/r, with a median of ~30% and going up to 40%. To increase robustness, the predictions are made by the consensus of all relevant models (weight by 1/MSE to account for the estimated differences in accuracy), as described in Supplement Methods. We do the same for the remaining four predictors, with the obtained median and maximal relative changes shown in Figure 6B. The obtained results confirm the importance of Chronic disease, Youth, Black, and Pollution, and, to a smaller extent, Population density.

**Discussion**

While we have earlier studied the parameters that might affect $R_0$ [19,21], the present goal was to investigate which demographic and environmental variables may influence the average disease severity as manifested in a population. The first step was to propose the response variable, which has to be causally independent of $R_0$ [19,21], to allow understanding the effects of clinical severity alone. We showed (both empirically and through the epidemiological model) that this is indeed satisfied by our choice (*m/r*). Additionally, this work allowed us to mechanistically interpret the standard (simple) measure of clinical severity (CFR), i.e., to relate its saturation value with the rate parameters in the epidemiological dynamical model. The relation is however non-linear (sigmoidal), which further underscores the non-triviality of the obtained result.

The proposed measure is practical to implement on a large scale (i.e., for diverse regions or countries, as we here demonstrated for 51 USA states), as only publicly available data are required, and calculation corresponds to a simple (though non-linear) relation. Estimating the saturation (end of the peak) is straightforward in most cases, through both case counts and *m/r* reaching a saturation (nearly constant) value. We set the following aims for the selected significant predictors of *m/r*: *i*) test if we can recover clinically observed dependencies, *ii*) uncover additional risk factors for COVID-19 clinical severity, suitable to extract from ecological study design [22], *iii*) compare with significant predictors of COVID-19 transmissibility ($R_0$) that we previously obtained [19,21]. We here indeed obtained different predictors for $R_0$ [19,21] compared to *m/r*. There are also some similarities consistent with inherent connections in COVID-19 transmissibility and severity drivers, e.g., the role of pollution, unhealthy living conditions, and indoor population density [19]. We further discuss *i*) and *ii*).

We obtain that both the prevalence of chronic diseases and Youth significantly influence *m/r*, which is hardly surprising - though quite a non-trivial result, as we started from a large group of initial variables. The influence of Disease PC4, dominantly reflecting the prevalence of cardiovascular diseases, is well documented by clinical studies [23,24] together with some other ecological studies [10,14]. Other chronic conditions that are well-known COVID-19 comorbidities (i.e., hypertension, obesity, and diabetes) are significant risk factors for cardiovascular diseases [25], and it is not surprising that cardiovascular diseases dominate over other chronic conditions in our results. Disease PC2, dominantly reflecting the prevalence of cancer (though also related to cardiovascular diseases), agrees with CDC warning that people with a history of cancer may be at increased risk of getting severely ill from COVID-19 [26]. Regarding Youth, it is well established that younger individuals are, on average, less severely affected by COVID-19, and that the disease severity increases with age [3,27,28].

We found that chronic pollution exposure, NOx levels in particular, significantly promote COVID-19 severity. While difficult to assess through clinical studies, it has been suggested that pollution is associated with the severity of COVID-19 conditions through similar pathways by which it affects respiratory and cardiovascular mortality [29]. In particular, NOx may reduce lung activity and increase infection in the airway [30]. Similarly, the effect of population density (which we found significantly affects *m/r*) is hardly suited to detect through clinical studies, while some ecological regression studies also noticed this dependence [31]. An explanation might be that while medical facilities are, in general, more abundant in overcrowded areas [32], this effect becomes overshadowed by the highly increased rate of the COVID-19 spread in these areas. Therefore, population density probably acts as a proxy for smaller health care capacity per infected (as the infections increase with the population density, particularly in indoor areas). Additionally, it was also proposed that higher viral inoculum may lead to more severe COVID-19 symptoms [33,34], where overcrowded conditions might lead to higher initial viral doses.

Another important predictor is the percentage of Afro-Americans (Black). While clinical evidence seems to confirm that Afro-Americans, in general, suffer from more severe COVID-19 symptoms (also obtained by several ecological meta-analyses [32,35], the underlying reasons are still a matter of debate (see e.g. [36]). Interestingly, this predictor appears only in non-parametric models, where interactions with other predictors are (implicitly) included. A posteriori, this result may not be surprising as it has been argued that higher clinical severity of COVID-19 for Black may be tightly related to other significant factors of COVID-19 severity (larger prevalence of chronic diseases, more crowded conditions, higher air pollution, which we here, in fact, obtain as significant predictors).

Finally, our rough estimates for the magnitude of the effects on *m/r* are also consistent with all four groups of factors (disease, youth/age, pollution, race) being significant drivers of COVID-19 severity, where a somewhat smaller magnitude was obtained for the fifth group (population density). Despite their significant association with $R_0$ [18,37]), meteorological variables were here not selected as significant predictors of COVID-19 clinical severity, which may be in part due to their smaller variability within the USA. On the other hand, we find that air pollution, which was previously also hypothesized to potentially contribute to COVID-19 severity [38], may indeed be a significant severity driver.

**Conclusion**

We employed a cross-disciplinary (One health) approach [39,40], combining epidemiological modeling with advanced statistical (machine) learning approaches, to explore the relationship of environmental factors to COVID-19 clinical severity. From an initially large number of variables (more than 60), we achieved a robust selection of a small number of significant factors, including those that are clinically known as determinants of COVID-19 severity. Our findings (performed in an unbiased manner directly from the data) are thus consistent with previous clinical studies. Additionally, our results underscore a syndemic nature of COVID-19 risks [41] through a selection of variables related to pollution, population density, and racial factors (intertwined with the effects of other factors). These results might have important implications for both longer and shorter-term efforts to alleviate the effects of this and (likely) future epidemics, in terms of longer-term policies to reduce these risks and shorter-term efforts to accordingly relocate medical resources. Our proposed measure (independent of disease transmissibility) originates from general considerations that are not limited to COVID-19. Thus, it may also be utilized in potential future outbreaks of infectious diseases, possibly also combined with other more traditional measures [42].

**Conflict of Interest**

The authors declare that the research was conducted in the absence of any commercial or financial relationships that could be construed as a potential conflict of interest.

**Supplementary Methods**

*Socio-demographic data collection*

Multiple sources were used for socio-demographic data. From the Social Science Research Council website [1] we obtained data on gender, race, population not between 18 and 65 (non-workforce), health insurance, infant and child mortality, life expectancy at birth and GDP. Center for Disease Control and Prevention (CDC) data [2] was reviewed to find medical parameters - cardiovascular disease, cholesterol, hypertension, inactivity, smoking, consuming alcohol, obesity, cancer, chronic kidney disease and chronic obstructive pulmonary disease (COPD). Census Reporter website [3] was used to obtain the percentage of the foreign population. Global Data Lab website [4] was used to obtain the Human Development Index (HDI) on the subnational level. U.S. Census Bureau website [5] was used to obtain the median age, population density, and urban population proportion.

*Pollution data collection*

US environmental protection agency (EPA) Air Data service [6] data was used to obtain air quality measures, which were aggregated on a daily level for all listed cities. Pollutant species monitored consisted of gases ($NO_2$, $CO$, $SO_2$, $O_3$), particulates ($PM_{2.5}$ and $PM_{10}$), Volatile Organic Compounds (VOC), $NO_x$ and Hazardous Air Pollutants (HAP). [7] was used to obtain populations of cities used for weighting the averages during aggregation. In further analysis, yearly averages for each pollutant species (representing chronic pollution exposure) were used.

*Weather data collection*

All the parameters pertaining to the weather were obtained from the NASA POWER project service [8] in an automated fashion using the POWER API and the custom Python scripts. Points of interest were coordinates obtained at Wikidata [9,10] for all the cities sorted by descending population size that comprises above 10% of the total country population. Parameters are listed in Table 1 and include temperature at 2m and 10m, measures of humidity and precipitation (wet bulb temperature, relative humidity, total precipitation), and insolation indices. The maximum daily predicted UV index was downloaded from OpenUV [11]. Weather parameters were then averaged for each USA state for the duration of the first peak.

*Data transformation and principal component analysis*

The distribution of most of the examined variables deviated from normality. To reduce skewness and the number of outliers in the data, appropriate transformations were applied (Table 1). Outliers are identified as values more than three scaled median absolute deviations (MAD) away from the median. After transformation, the remaining outliers were substituted by the transformed variable median value.

To reduce the number of variables, which was initially larger (62) than the sample size (51), we divided data into mutually related subsets and performed Principal Component Analysis (PCA) [12], on each group (Table 2). Grouping of the variables was done following two criteria: *i)* variables present similar quantities to allow for easier interpretation of the principal components, *ii)* they are highly mutually correlated. Consequently, after PCA, correlations between the resulting predictors are reduced. Variables that did not satisfy the above criteria were not grouped, and they were used in the analysis as they are. Additionally, the variables that contributed to the relevant PCs in a way that was hard to interpret, were also treated as independent predictors. For example, the percentage of the youth population (Youth) has an opposite meaning from the other two age-related variables (Median age and percent of the population over 65), so that it was treated as an independent variable.

The number of PCs retained for each group was determined to explain >85% of the data variance. Afterward, a total of 29 variables (18 principal components and 11 independent variables) remained.

| Data | Name (units) | Transformation f(x) |
|---|---|---|
| m/h | Morbidity | $x^{1/2}$ |
| T2M | The mean temperature at 2m (°C) | $(x - (x))^{1/3}$ |
| T2M$_{MAX}$ | The average maximal temperature at 2 meters (°C) | None |
| T2M$_{MIN}$ | The average minimal temperature at 2 meters (°C) | $(x - (x))^{1/3}$ |
| T10M | The mean temperature at 10 meters | $x^{1/3}$ |
| T10M$_{MAX}$ | The average maximal temperature at 10 meters (°C) | None |
| T10M$_{MIN}$ | The average minimal temperature at 10 meters (°C) | $(x - (x))^{1/3}$ |
| TS | The surface temperature (°C) | $log(x - min(x))$ |
| T2MWET | The wet-bulb temperature at 2 meters (°C) | None |
| RH2M | Relative humidity at 2 meters (%) | $-log(max(x) - x)$ |
| QV2M | Specific humidity at 2 meters (g/kg) | $log(x)$ |
| T2MDEW | Dew point (°C) | None |
| PRECTOT | Precipitation (mm/day) | $x^{1/3}$ |
| TQV | Total Column Precipitable Water (cm) | $log(x)$ |
| UV | UV radiation index | $x^{1/2}$ |
| ALLSKY_SFC_SW_DWN | All Sky Insolation Incident on a Horizontal Surface (MJ/m$^2$/day) | None |
| CLRSKY_SFC_SW_DWN | Clear Sky Insolation Incident on a Horizontal Surface (MJ/m$^2$/day) | $-log(max(x) - x)$ |
| ALLSKY_SFC_LW_DWN | Downward Thermal Infrared (Longwave) Radiative Flux (MJ/m$^2$/day) | $log(x)$ |
| PS | Pressure (mbar) | $-log(max(x) - x)$ |
| WS2M | Wind speed at 2 meters (kts) | $-log(max(x) - x)$ |
| WS10M | Wind speed at 10 meters (kts) | $x^2$ |
| Elderly | Population over 65 (%) | None |
| Median age | Median age (years) | $log(x)$ |
| Youth | Population under 18 (%) | $-((x) - x)^{1/2}$ |
| Population density | Population density (people/km$^2$) | $log(x)$ |
| BUAPC | Built-Up Area Per Capita (km$^2$/people) | $x^{-1/3}$ |
| Urban Population | Urban population (%) | $x^2$ |
| GDPpc | Gross Domestic Product per capita | $log(x)$ |
| HDI | Human Development Index | $-((x) - x)^{1/2}$ |
| Infant mortality | Infant mortality rate (per 1,000 live births) | $-log(x)$ |
| Child mortality | Child mortality rate (per 1,000 live births) | $-log(x)$ |
| Alcohol consumption | Adults' alcohol consumption binge drinking (%) | $log(x)$ |
| Foreign-born population | Foreign-born population (%) | $log(x)$ |
| Life expectancy | Life expectancy at birth (years) | $-((x) - x)^{1/2}$ |
| Obesity | Obesity at age 20 and older (%) | None |
| CVD deaths | Age 65+ cardiovascular disease deaths (per 100,000 people) | $log(x)$ |
| Hypertension | Adults with hypertension (%) | $log(x)$ |
| High cholesterol | Population with high cholesterol (%) | $x^{1/2}$ |
| Smoking | Population smoking (%) | None |
| Cardiovascular disease | Population with cardiovascular disease (%) | None |
| Diabetes | Population with diabetes (%) | $x^{1/2}$ |
| Cancer | Population with cancer (%) | None |
| CKD | Population with chronic kidney disease (%) | $x^{1/3}$ |
| COPD | Population with chronic obstructive pulmonary disease (%) | $log(x)$ |
| Multiple chronic conditions | Population with multiple chronic conditions (%) | $log(x)$ |
| Physical Inactivity | Physically inactive population (%) | $x^{1/2}$ |
| White | Fraction of white in the population (%) | $x^2$ |
| Black | Fraction of Afro-Americans in the population (%) | $x^{1/3}$ |
| Latino | Fraction of Latino in the population (%) | $log(x)$ |
| No Insurance Children | No health insurance under 18 (%) | $x^{1/2}$ |
| No Insurance Adults | No health insurance 18-64 (%) | $x^{1/2}$ |
| No Insurance Total | No health insurance all population (%) | $x^{1/2}$ |
| No Insurance Black | No health insurance among black (%) | None |
| No Insurance Latino | No health insurance Latino (%) | None |
| No Insurance White | No health insurance white (%) | None |
| $PM_{2.5}$ | $PM_{2.5}$ concentration (µg/m$^3$) | $-\left((x)-x\right)^{1/2}$ |
| $PM_{10}$ | $PM_{10}$ concentration (µg/m$^3$) | $x^{1/3}$ |
| $O_3$ | $O_3$ concentration (ppm) | $x^2$ |
| CO | CO concentration (ppm, $10^{-6}$) | None |
| $SO_2$ | $SO_2$ concentration (ppb) | $x^{1/3}$ |
| HAPs | Hazardous air pollutants concentration (µg/m$^3$) | $x^{1/3}$ |
| $NO_2$ | $NO_2$ concentration (ppb, $10^{-9}$) | $x^{1/3}$ |
| NONOxNOy | Nitrous oxide concentration (ppb) | $x^{1/2}$ |

**Table 1:** List of variables with appropriate transformations.

| Principal components | Variables |
|---|---|
| Temperature PC1 | T2M, T2M$_{MAX}$, T2M$_{MIN}$, T10M, T10M$_{MAX}$, T10M$_{MIN}$, TS, T2MWET |
| Humidity PC1 | RH2M, QV2M, T2MDEW |
| Precipitation PC1 | PRECTOT, TQV |
| Radiation PC1-PC2 | ALLSKY_SFC_SW_DWN, CLRSKY_SFC_SW_DWN, ALLSKY_SFC_LW_DWN |
| Wind Speed PC1 | WS2M, WS10M |
| Seasonality PC1-PC2 | Temperature PC1, Humidity PC1, Precipitation PC1, UV, Radiation PC2 |
| Age PC1 | Elderly, Median age |
| Density PC1-PC2 | Population density, BUAPC, Urban population |
| Prosperity PC1-PC4 | GDPpc, HDI, Infant Mortality, Child mortality, Alcohol consumption, Foreign-born, Life expectancy |
| Disease PC1-PC4 | Obesity, CVD deaths, Hypertension, High Cholesterol, Smoking, Cardiovascular disease, Diabetes, Cancer, CKD, COPD, Multiple chronic conditions, Physical inactivity |
| No Insurance PC1-PC2 | No Insurance Children, No Insurance Adults, No Insurance Total, No Insurance Black, No Insurance Latino, No Insurance White |
| NO PC1 | $NO_2$, NONOxNOy |

**Table 2:** Grouping of the variables before PCA and selected principal components from each group.

*Relaxed LASSO regression*

A modification of Lasso (Least Absolute Shrinkage and Selection Operator) [13] regression, Relaxed Lasso [14], was used to implement L1 regularization on high-dimensional data. Selected 29 variables were standardized before the first Lasso regression analysis. Hyperparameter λ was optimized by 5-fold cross-validation, with 40 dataset repartitions. 100 λ values in the range from 0 to the minimal λ value (which produces all zero terms) were put on the grid, where the optimal λ value was determined as having minimal MSE (Mean Squared Error) on the testing set. This hyperparameter value was used to train the first round model on the entire dataset. Only predictors with non-zero coefficients from this model were used in the second (relaxed) Lasso regression. The optimal λ value in the second round was determined by cross-validation as described for the first round, which was then used to train the final (second round) model on the entire dataset. By using Relaxed Lasso regression, noise from the high-dimensional data (in particular those variables that do not influence the output) is reduced, allowing for more accurate estimates of the reported regression coefficients. The final model from the second round was used for subsequent predictions, with its regression coefficients reported. This, and the other three procedures described below, were trained both with and without No Insurance (% of the uninsured population) data.

*Relaxed Elastic net regression*

Elastic net regression [15] was used for the implementation of L1 and L2 regularization. The procedure was similar to the Relaxed Lasso analysis explained above, only this time two hyperparameters – α and λ were optimized. These hyperparameters were put on a grid consisting of 100 uniformly distributed (from 0 to 1) α values, and 100 λ values chosen for each α value as described for the Lasso regression. Similarly, as for Lasso, 5-fold cross-validation with 40 dataset repartitions was used. Optimal α and λ values were determined as those with minimal testing set MSE, which were used to train the first round model on the entire dataset. Predictors with non-zero coefficients from the final first-round model were used as an input for the second (relaxed) Elastic net round. Optimal hyperparameter (α and λ) values were determined by cross-validation equivalently as in the first round, which were then used to train the final (second round) model on the entire dataset. Regression coefficients obtained from the final model were reported, which was subsequently also used for predictions.

*Random Forest and Gradient Boost*

Ensembles of weak learners (decision trees) were implemented through Random Forest and Gradient Boost [16–19]. Optimal hyperparameters were determined by grid search, with 5-fold cross-validation and 40 dataset repartitions, equivalently to Lasso and Elastic net regressions. In each cross-validation round, input variables were preselected based on their significant correlations ($P < 0.1$ for either Pierson's, Spearman, or Kendall) with $m/r$ on the testing set. This is to avoid overfitting by reducing the number of model predictors. For Random Forest, maximal number of splits, minimal leaf size and number of trained decision trees on the grid were respectively: {3, 6, 9, 12, 18, 20, 22, 24, 26, 30, 35, 37, 40, 43, 45, 47, 50}, {1, 2, …, 10}, {5, 8, 10, 14, 18, 26, 50, 106, 133, 160, 193, 226, 263, 300, 350, 400, 450, 500, 550, 600}. For Gradient Boost values of maximal number of splits, minimal leaf size, number of trained decision trees and learn rate on the grid were respectively: {1, 2, 3, 4, 5, 6, 7, 8, 10, 12, 14, 16, 24, 32}, {1, 2, 3, 4, 5, 8, 12, 16, 18, 20, 22, 25}, {5, 10, 13, 16, 21, 27, 38, 65, 92}, {0.1, 0.20, 0.25, 0.35, 0.5, 0.75, 1}. Combinations of the hyperparameter values that lead to the minimal testing set MSE were used to train the final models on the whole dataset. The input variable preselection in the final models was done on the entire dataset, equivalently as described above. Final models were used to estimate the predictor importance and in the predictions described below.

*Predictions of δ(m/r)*

Regression predictions of *δ(m/r)* were made by consensus, i.e., averaging the following final models described above: *i*) For chronic disease, population density, and pollution, all eight models were used (Lasso, Elastic net, Random Forest, Gradient Boost, each trained both with and without No Insurance). While $R^2$ for the decision tree based methods (Random Forest and Gradient Boost) is larger than for the linear regressions, the differences in the testing set MSE (prediction accuracy) were not large, so all eight models were used to achieve robust results. *ii*) For the percentage of population under 18 (Youth) the four models trained without No Insurance were used, as the strong correlation between Youth and No Insurance obscures the relation of Youth to *m/r*. *iii*) For the percentage of Afro-Americans (Black), the four nonparametric models (that can accommodate non-linear relations and interactions) were used (Gradient Boost and Random Forest both with and without No Insurance), as the contribution of this variable to *m/r* is not captured by linear regressions. All the averages above are weighted by 1/MSE so that models with higher prediction accuracy are included with larger weights.

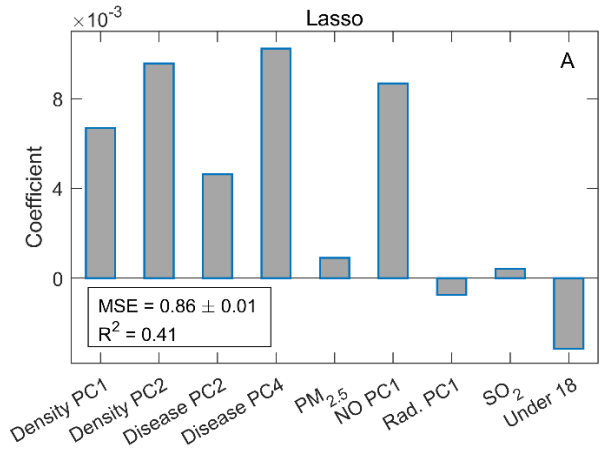
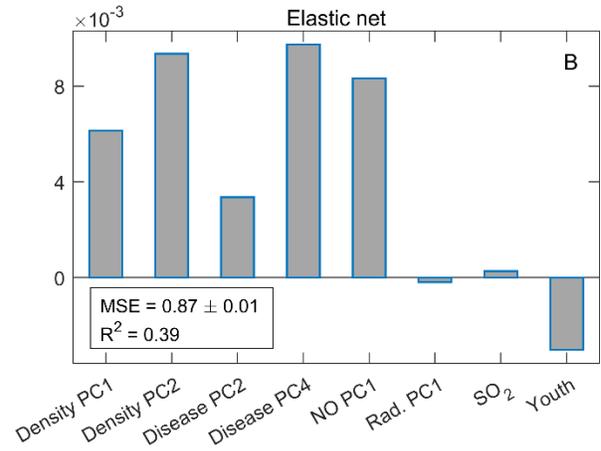
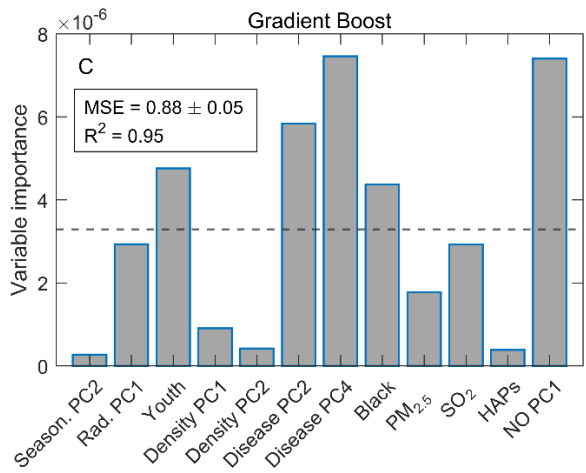
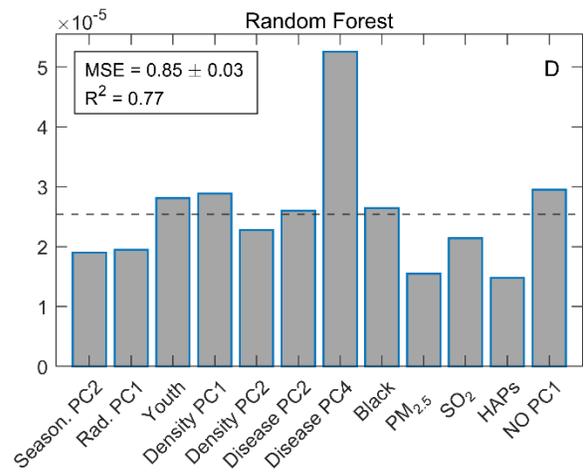

**Figure S1: Multivariate (machine learning) analysis without No insurance PC.** The equivalent analysis as in Figure 5 is performed, but with excluded No Insurance PC data. The explanation is the same as in the caption of Figure 5.